# 基于卡诺机的大规模 MIMO 通信容量建模及性能分析

阮慕瑶[1,2]，葛晓虎[1,2,*]

（1. 华中科技大学 电子信息与通信学院，武汉 430074;2. 国家绿色通信与网络国际联合研究中心，武汉 430074）

**摘　要：**信息本质上以能量和熵的形式在通信系统中流动，类似于传统热机中的能量流动过程。本文结合热力学卡诺机模型和经典香农信息论理论，建立了一个广义的热力学 MIMO(multiple input multiple outputs , MIMO)通信系统模型，对大规模天线通信场景下使用前向纠错码的信道容量进行分析。利用通信系统中存在的自由度和熵的概念，推导了热力学模型下的广义信道容量，并对信道容量与噪声自由度及编码开销的关系进行验证仿真。仿真结果表明本文提出的热力学信道容量与香农信道容量是一致的。

# Carnot machine-based massive MIMO communication capacity modeling and performance analysis

RUAN Muyuo[1,2], GE Xiaohu[1,2,*]

（1. School of Electronic Information and Communications, Huazhong University of Science and Technology, Wuhan 430074, China; 2. the International Joint Research Center of Green Communications and Networking, Wuhan 430074, China）

**Abstract:** Similar to the energy flowing process in traditional heat engines, information could be considered to flow in the communication systems with the form of energy and entropy. Combining the thermodynamic Carnot machine and the classical Shannon information theory, a generalized thermodynamic MIMO (multiple input multiple outputs) communication system is established to analyze the channel capacity using forward error correction codes. Based on the concepts of freedom and entropy in the communication system, the generalized channel capacity is proposed under the thermodynamic theory. Furthermore, the relationships between the proposed channel capacity and the noise freedom and coding overhead are derived and simulated. Simulation results verify the proposed channel capacity is coincident with the classical channel capacity.

**收稿日期：**2021-12-20　**修订日期：**2022-02　**通讯作者：**葛晓虎　xhge@hust.edu.cn

**基金项目：**国家自然科学基金(62120106007)

Foundation Item: The National Natural Science Foundation of China (62120106007)

## 0　引　言

信息论自从诞生以来就和热力学有着密不可分的关系。1948 年香农开创性地提出信息熵这一概念，实际上也是从热力学中熵的概念引申而来。信息熵用来描述信源的不确定度，热力学中的热力学熵则表示分子状态混乱程度的物理量，如冯诺依曼熵描述的是物理系统中粒子不同的状态。如今信息论在热力学的多分支领域均有应用[1]，另一方面也带来了如何将热力学应用到通信系统中的思考。

一直以来，人们认为在非理想环境下通信过程一定伴随着能量的耗散。物理系统中发射端通过通信信道和接收端相连，被编码的粒子状态信息在信道中的传输被称之为信息传输[2]，同样需要能量的耗散。现实中非理想热机由于不可逆的耗散不能达

到卡诺热机效率极限。因此，如何在有限的条件下传输信息而使耗散能量最低成为众多领域的研究热点。Ganesh 等人用物理分析方法提出了前馈神经网络耗散的基本下界[3]。Victoria 等人研究了在无记忆高斯信道中在一定错误率的限制下传输指定符号数所需最小传输能量[4]。这引发了进一步的思考，即现实通信系统中的热力学物理限制和信息传输效率之间的内在关系。

卡诺热机定理给出了热力学约束下的热机最大工作效率$\eta = 1 - T_{\mathrm{LO}} / T_{\mathrm{HI}}$，其中$T_{\mathrm{HI}}$和$T_{\mathrm{LO}}$分别是高温热源温度和低温热源温度。香农定理给出了一定约束条件下的物理信道最大传输数据速率，卡诺热机效率衡量了热能转化为有用功的最大效率，其背后原理是热力学定律。高斯信道可以被看作一个被热力学定律所约束的物理信息系统[5],[6]，并且已有研究表明信道容量可以被重新表述为热力学定律的结果，其中对高斯噪声信道和二进制对称信道的推论都进行了例证[7]。通过热力学这一桥梁，卡诺热机定理和香农信道定理展示出本质上的共通性[8]。近年来在信息热力学领域，将信息论和热机概念结合分析通信系统的研究逐渐涌现。Shental 等人运用哈密顿量将通信信道建模为信息热机，讨论高斯信道的互信息等特性[9]。Parker 等人利用经典卡诺定理分析了兰道尔原理和香农极限的一致性，并推导了包含非高斯噪声统计量的广义香农信道容量定理[10]。Chen 等人提出的一个广义不可逆卡诺热机模型分析了最大输出功率和最大效率之间的关系[11]。

此外，大多数文章只是建立在端到端的通信系统模型上分析，着重于对广义通信和热力学之间关系的阐述与推导[9]-[11]，而未基于现有的多天线通信系统场景下结合热力学对信道容量进行分析推导，从而进一步验证通信和热力学的内在关联。本文更加详细地研究 5G（5th Generation, 5G）大规模 MIMO（multiple input multiple outputs，MIMO）多天线场景[12],[13]，从分析能量流动的普遍适用工具——热力学第二定律出发，以全新的物理信息角度分析了大规模 MIMO 通信系统信息流的传输和处理过程。结合卡诺热机定理和信息论，引入自由度、温度、能量、熵等热力学参数构建了大规模 MIMO 场景下的信道容量，揭示了无线通信系统中信息、熵与能量之间的本征关系。

本文的结构如下，第一章构建了基于热力学的 MIMO 通信系统。第二章推导了基于热力学的 MIMO 信道容量，第三章对所提出的信道容量进行了仿真分析，展示了热力学信道容量和各参量之间变化关系，对比了基于热力学的广义信道容量与香农信道容量的差异，并给出了解释。

# 1 系统模型

## 1.1 信息热机中能量的流动形式

热力学理论，特别是热力学第二定律，是分析能量流动的普遍适用工具[14]。本章将介绍信息热机和传统卡诺机的相似性，在卡诺热机模型基础上结合热力学理论对 MIMO 通信系统中的能量流和信息流进行建模分析。

如图 1（a），热机进行卡诺循环从一个高温热库$T_{HI}$吸取热量$Q_H$，转化为一部分功$W$，剩余热量$Q_L = Q_H - W$排入低温热库$T_{\mathrm{LO}}$。图 1（b）中当熵被视作为一种做功潜力的度量时[10]，低熵代表系统更加有序[15]，有序系统比无序系统的做功潜力更高。低熵库的熵流入热机被转换为一部分功，剩余熵流入高熵库，本文中熵流被表示为热量与相应温度的比值，如图 1（b）中的$\Delta H_1$和$\Delta H_2$。

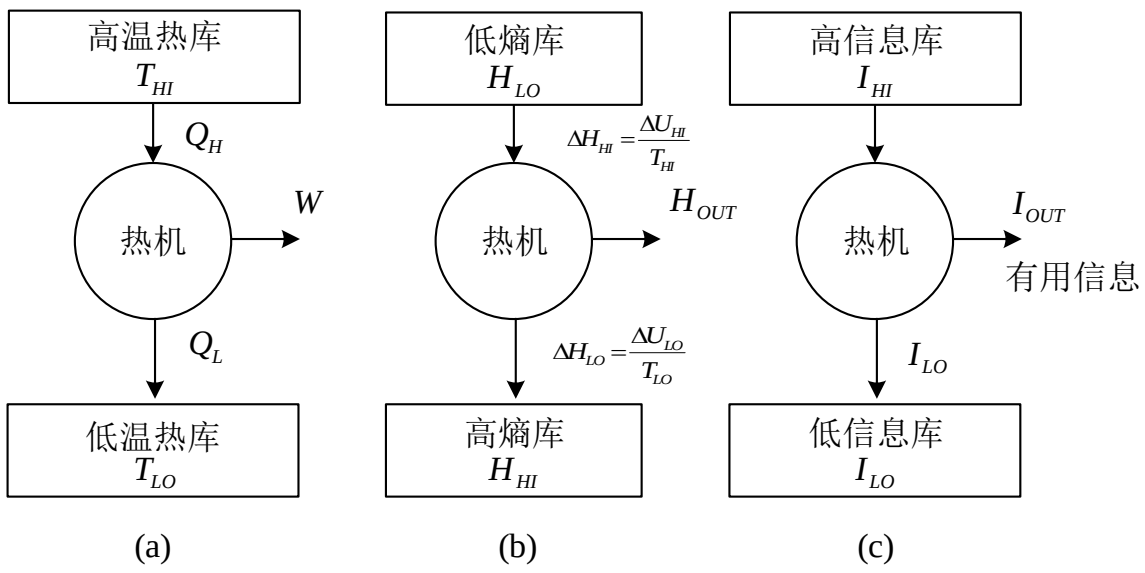

**图 1（a）传统热力学热机原理图 （b）等效的熵热机原理图 （c）等效信息热机原理图**

Fig.1 (a) Schematic diagram of traditional thermodynamic heat engine (b) Schematic diagram of equivalent entropy heat engine (c) Schematic diagram of equivalent information heat engine

图 1（c）是信息热机原理图，热机可以对应通信过程中的信息处理模块甚至是信道，热机从高信息库获得的信息转化为一部分输出功，即有用信息$I_{\mathrm{OUT}}$，剩余的被排入低信息库，例如现实中的算法对输入数据进行处理后输出想要的计算结果，但是总会丢弃一部分冗余或者不需要的额外信息（变量）。我们将信息热机的效率定义为$\eta = I_{\mathrm{OUT}} / I_{\mathrm{HI}}$。接下来将根据以上信息热机模型构建一个具体的大规模 MIMO 通信系统，进而对各部分进行详细分析。在此之前，我们认为有必要介绍一下信息的物理表达形式。

## 1.2 信息的物理表达形式

系统的宏观状态以微观状态的分布为特征，这个分布的熵由 Gibbs 熵公式给出。对于一串由脉冲组成的信息序列，不同码字对应着不同微观态。在二进制场景下对应的微观态表示为 0 或 1。则单个码字的 Gibbs 混合熵为[16]

$$H_i = -k_{\mathrm{B}} \sum_{j=1}^{\Omega} p_j \ln p_j \tag{1}$$

(1)式中：$\Omega$是信息码元种类集合数，$k_{\mathrm{B}}$是玻尔兹曼常数（$k_{\mathrm{B}}=1.38\times10^{-23}$）。$p_j$为每种状态出现的概率。假设一串长度为$M$比特的二进制序列，其中符号之间相互独立且服从$\Pr\{X=1\}=\Pr\{X=0\}=1/2$的伯努利分布，$\Pr$表示概率。则序列的 Gibbs 总熵为：

$$H=\sum_{i=1}^{M}H_i=\sum_{i=1}^{M}\left(-k_{\mathrm{B}}\left(\frac{1}{2}\ln\frac{1}{2}+\frac{1}{2}\ln\frac{1}{2}\right)\right)=k_{\mathrm{B}}M\ln 2 \tag{2}$$

在众多领域中，熵是一种度量系统不确定性状态的量，同时熵也可以度量系统的自由度数量[17]。信息序列的熵越大，表示它所携带的信息量越大，系统的自由度也更大。例如，1 bit 指一位二进制符号所携带的信息量，或者是从 0, 1 两种状态选择其一的自由度，也是下一时刻选择何种符号发送的自由度。在这里我们引入自由度这一概念来表征信息序列。根据(2)式，信息序列的 Gibbs 熵$H$和自由度$M$之间的关系为：

$$H=k_{\mathrm{B}}M\ln 2 \tag{3}$$

根据热力学温度的表达式$\frac{1}{T}=\frac{dH}{dU}$[10]，[18]，能量$U$，Gibbs 熵$H$（以下涉及“熵”的推导中均表示“Gibbs 熵”的简述）与自由度$M$之间的关系为

$$H=\frac{U}{T}=k_{\mathrm{B}}M\ln 2 \tag{4}$$

### 1.3 大规模 MIMO 通信系统模型

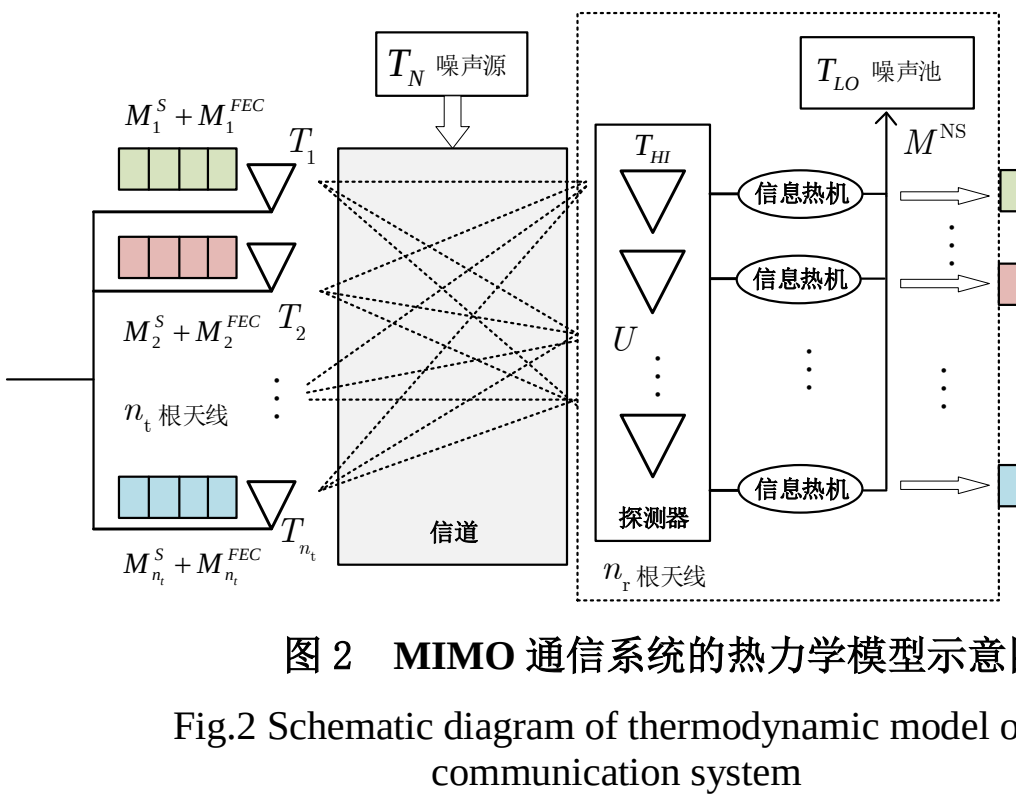

**图 2　MIMO 通信系统的热力学模型示意图**

Fig.2 Schematic diagram of thermodynamic model of MIMO communication system

如图 2 所示，这是一个$n_{\mathrm{t}}\times n_{\mathrm{r}}$的大规模 MIMO 通信系统，$n_{\mathrm{t}}$为发射端天线数，$n_{\mathrm{r}}$为接收端天线数，通常在大规模 MIMO 通信场景下$n_{\mathrm{t}}\gg n_{\mathrm{r}}$。每条支路上原始自由度为$M^{\mathrm{S}}$的信息序列增加冗余检错码后被映射为自由度为$M^{\mathrm{S}}+M^{\mathrm{FEC}}$的码字（$M^{\mathrm{FEC}}$为冗余纠错码的自由度），再经由多根天线发送出去。信道噪声将每个码字随机地转换为多个序列的统计混合，然后通过接收端的译码操作将每个序列映射成原始信息中的对应码字。在这里冗余纠错位为前向纠错码（Forward Error Correction，FEC），前向纠错码通过在已编码的消息中构建冗余来提供噪声抗扰性。在接收端通过不可逆转(即逻辑上不可逆)解码操作将冗余从信道损坏的码字中移除从而大概率地恢复出原始信息，剩余的冗余信息将流入图 2 中的噪声池。这种逻辑上的不可逆解码操作将冗余从信道损坏的码字中移除的过程，对应的物理实现过程也会不可避免地导致热量耗散。正如兰道尔提出的，“解码本质上是能量耗散的”[19]，[20]。

令通信系统各部分由参量集合$\{U,M,H,T\}$表示，$U$是能量，$M$是自由度，$H$是熵，$T$是温度。这些参量之间的关系如(4)式所示。其中自由度满足可加性原则，例如源编码序列自由度$M^{\mathrm{S}}$加上前向纠错码自由度$M^{\mathrm{FEC}}$之后的自由度表示为$M^{\mathrm{S}}+M^{\mathrm{FEC}}$。经由$n_{\mathrm{t}}$根发射天线的总自由度$M_{\mathrm{send}}^{\mathrm{tot}}$表示为

$$\begin{aligned}M_{\mathrm{send}}^{\mathrm{tot}}&=\frac{U_1+U_1^{\mathrm{FEC}}}{T_1}+\frac{U_2+U_2^{\mathrm{FEC}}}{T_2}+\dots+\frac{U_{n_{\mathrm{t}}}+U_{n_{\mathrm{t}}}^{\mathrm{FEC}}}{T_{n_{\mathrm{t}}}}\\&=\sum_{i=1}^{n_{\mathrm{t}}}\left(\frac{U_i+U_i^{\mathrm{FEC}}}{T_i}\right)\end{aligned} \tag{5}$$

有噪信道中引入的各支路噪声自由度之和为

$$M_{\mathrm{N}}^{\mathrm{tot}}=\sum_{j=1}^{n_{\mathrm{r}}}\left(\frac{U_j^{\mathrm{N}}}{T_j^{\mathrm{N}}}\right) \tag{6}$$

(5)(6)式中，上标N表示与噪声有关的参量，上标FEC表示与前向纠错码有关的参量。下标$i$ $j$ 表示第$i$ $j$ 根发射（接收）天线端信号有关的参量，$T_i$为第$i$根天线发射信号的温度，$U_i$为第$i$根天线发射信号的能量，$1\le i\le n_{\mathrm{t}}$，$1\le j\le n_{\mathrm{r}}$。

假设整个信息过程中的自由度守恒，即经由所有发送天线的自由度与信道中引入的噪声自由度之和等于到达接收端探测器的所有自由度。由(4)式，探测器端的自由度表示为

$$M_{\mathrm{dect}}=\frac{U}{T_{\mathrm{HI}}} \tag{7}$$

(7)式中：$U$表示接收端探测器接收到的总能量，$T_{\mathrm{HI}}$为探测器的温度。结合(5)(6)(7)式

$$\frac{U}{T_{\mathrm{HI}}}=\sum_{i=1}^{n_{\mathrm{t}}}\left(\frac{U_i+U_i^{\mathrm{FEC}}}{T_i}\right)+\sum_{j=1}^{n_{\mathrm{r}}}\left(\frac{U_j^{\mathrm{N}}}{T_j^{\mathrm{N}}}\right) \tag{8}$$

(8)式中的$T_{\mathrm{HI}}$可进一步表示成

$$T_{\mathrm{HI}}=\frac{\left(\prod_{i=1}^{n_{\mathrm{t}}}T_i\right)\times U}{\sum_{i=1}^{n_{\mathrm{t}}}\left[\left(\prod_{\substack{j=1\\ j\neq i}}^{n_{\mathrm{t}}}T_j\right)\times U_j+U_j^{\mathrm{FEC}}\right]+\frac{\left(\prod_{i=1}^{n_{\mathrm{t}}}T_i\right)}{\prod_{j=1}^{n_{\mathrm{r}}}T_j^{\mathrm{N}}}\times\sum_{p=1}^{n_{\mathrm{r}}}\left[U_p^{\mathrm{N}}\times\prod_{\substack{q=1\\ q\neq p}}^{n_{\mathrm{r}}}T_q^{\mathrm{N}}\right]} \tag{9}$$

当发射端天线远大于接收天线数时 $n_{\mathrm{t}}\gg n_{\mathrm{r}}$ (9)式中分母的第二项可以近似为 0。也可以说在这种极限情况下，噪声的影响可以忽略不计。当各支路信噪很小时，即 $U_j^{\mathrm{N}}\gg U_i+U_i^{\mathrm{FEC}}$，$T_{\mathrm{HI}}\to\left(\prod_{j=1}^{n_{\mathrm{r}}}T_j^{\mathrm{N}}\right)/\left(\sum_{j=1}^{n_{\mathrm{r}}}T_j^{\mathrm{N}}\right)$。当各支路信噪比很大时，即 $U_j^{\mathrm{N}}\ll U_i+U_i^{\mathrm{FEC}}$，$T_{\mathrm{HI}}\to\left(\prod_{i=1}^{n_{\mathrm{t}}}T_i\right)/\left(\sum_{i=1}^{n_{\mathrm{t}}}\alpha_i T_i\right)$，其中 $\alpha_i$ 为归一化系数，$\sum_{i=1}^{n_{\mathrm{t}}}\alpha_i=1$。

### 1.4 传输每比特需要耗散的能量

对于图 2 中虚线框内的解码模块，探测器（如天线系统）将接收到的所有支路信号经由信息处理器（可理解为译码）恢复出原始信号流，从信道中携带的一部分噪声以及最初的前向纠错码所携带的自由度流入噪声池。如图 3 所示。

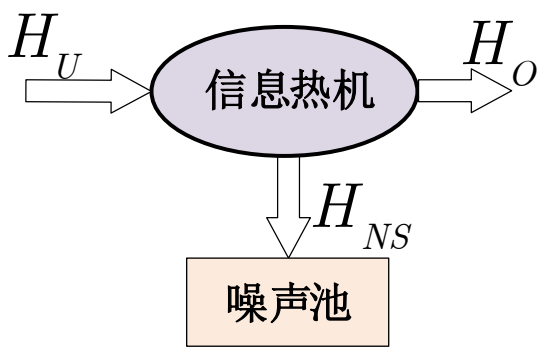

**图 3 接收端进行信息译码操作的熵变过程**

Fig.3 The entropy evolution during information decoding operation

在这个过程中，熵是守恒的，因此

$$H_{\mathrm{U}}=H_{\mathrm{O}}+H_{\mathrm{NS}} \tag{10}$$

(10)式中：$H_{\mathrm{U}}$ 是接收端所有接收信号的熵，$H_{\mathrm{O}}$ 是经过解码模块后（对应图 3 中的信息热机）的信息序列所携带的熵，$H_{\mathrm{NS}}$ 是流入噪声池（noise sink，NS）的熵。由式(4)(10)得：

$$\frac{U}{T_{\mathrm{HI}}}=\sum_{j=1}^{n_{\mathrm{r}}}\left(\frac{U_j^{\mathrm{O}}}{-T_j^{\mathrm{O}}}\right)+\frac{U-\sum_{j=1}^{n_{\mathrm{r}}}U_j^{\mathrm{O}}}{T_{\mathrm{LO}}} \tag{11}$$

(11)式中：$U_j^{\mathrm{O}}$ 和 $T_j^{\mathrm{O}}$ 分别是第 $j$ 条解码支路的信号能量和温度，$T_{\mathrm{LO}}$ 是噪声池温度。

假设发射端每根天线发送的信号自由度相同，将传输每比特所需要耗散的能量定义为流入噪声池的总热量 $U-\sum_{j=1}^{n_{\mathrm{r}}}U_j^{\mathrm{O}}$ 与源信息序列的总自由度 $\sum_{j=1}^{n_{\mathrm{r}}}M_j^{\mathrm{O}}$ 的比值。则传输每比特所需要耗散的能量表示为

$$E_{\mathrm{per_bit}}=\frac{U-\sum_{j=1}^{n_{\mathrm{r}}}U_j^{\mathrm{O}}}{\sum_{j=1}^{n_{\mathrm{r}}}M_j^{\mathrm{O}}}=\frac{1+\dfrac{\sum_{j=1}^{n_{\mathrm{r}}}T_j^{\mathrm{O}}}{n_{\mathrm{r}}\times T_{\mathrm{HI}}}}{1-\dfrac{T_{\mathrm{LO}}}{T_{\mathrm{HI}}}}k_{\mathrm{B}}T_{\mathrm{LO}}\ln 2 \geq k_{\mathrm{B}}T_{\mathrm{LO}}\ln 2 \tag{12}$$

(12)式中：$M_j^{\mathrm{O}}$ 为第 $j$ 条支路解码得到的信号自由度。由(12)式可以看到 $E_{\mathrm{per_bit}}$ 随着噪声温度池 $T_{\mathrm{LO}}$ 的升高而升高。当 $T_{\mathrm{HI}}\geq T_j^{\mathrm{O}},T_{\mathrm{HI}}\geq T_{\mathrm{LO}}$ 时：

$$\min\ E_{\mathrm{per_bit}}=k_{\mathrm{B}}T_{\mathrm{LO}}\ln 2 \tag{13}$$

这表示传输 1bit 所需要耗散的最小热量为 $k_{\mathrm{B}}T_{\mathrm{LO}}\ln 2$，这也是 $E_{\mathrm{per_bit}}$ 的下界。这个结论在一定程度上非常符合兰道尔提出的“擦除 1 bit 信息所需的能耗下限为 $k_{\mathrm{B}}T\ln 2$”，即兰道尔原理[21]。对于流入噪声池中的残余信息流

$$\sum_{j=1}^{n_{\mathrm{r}}}M_j^{\mathrm{NS}}=\frac{U-\sum_{j=1}^{n_{\mathrm{r}}}U_j^{\mathrm{O}}}{T_{\mathrm{LO}}} \tag{14}$$

(14)式中：$\sum_{j=1}^{n_{\mathrm{r}}}M_j^{\mathrm{NS}}$ 为流入噪声池的总自由度。进而由式(12)(14)得

$$\sum_{j=1}^{n_{\mathrm{r}}}M_j^{\mathrm{O}}=\eta_c\sum_{j=1}^{n_{\mathrm{r}}}M_j^{\mathrm{NS}} \tag{15}$$

(15)式中：卡诺效率 $\eta_c=1-\dfrac{T_{\mathrm{LO}}}{T_{\mathrm{HI}}}$。

(15)式表明了最终得到的信号自由度和流入噪声池的所有自由度之间的比例关系，而系数为卡诺效率 $\eta_c$。当高温热池 $T_{\mathrm{HI}}$ 和低温噪声池 $T_{\mathrm{LO}}$ 之间的温度差 $dT$ 越大，卡诺效率越高，实际最终得到的信息量也越多，即信息传输效率越高。

## 2 信道容量模型

利用信道矩阵数学方法计算出 $m\times m$ MIMO 通信系统的香农信道容量为

$$C = B \times \sum_{i=1}^{m} \log_2 \ 1 + \lambda_i^2 \rho_i \qquad (16)$$

(16)式中：$B$ 为信道带宽，$\rho_i$ 为第 $i$ 个子信道的信噪比，$\lambda_i$ 为第 $i$ 个子信道的功率增益。接下来我们将利用热力学参量推导出广义热力学信道容量，并说明其与(16)式的不同之处。

假设噪声功率定义为 $P_{\rm N} = k_B TB$，$T$ 为环境温度。则单位时间 $\tau$ 内流入噪声池内的能量 $U_{\rm NS} = P_{\rm N}\tau = k_{\rm B} M^{\rm NS} T_{\rm LO} \ln 2$，进一步得到在接收端解码过程的第 $j$ 条支路流入噪声池的自由度近似为

$$M_j^{\rm NS} = \frac{k_{\rm B} B}{\ln 2} \qquad (17)$$

代入(15)式并对等式两边同时微分

$$dM_j^{\rm O} = \frac{\tau B}{\ln 2}\frac{1}{T_j^{\rm IP}} dT_j^{\rm IP} \qquad (18)$$

(18)式中：$T_j^{\rm IP}$ 为解码端的信号温度，定义微分形式的卡诺效率 $\eta_c = \dfrac{dT}{T}$。基于(18)式将接收端所有支路最终解码出的信息自由度进行求和积分

$$\int_0^{\sum_{j=1}^{n_{\rm r}} M_j^{\rm S}} dM_{\rm S} = \frac{\tau B}{\ln 2} \int_{T_1^{\rm N}}^{T_1^{\rm IP}} \int_{T_2^{\rm N}}^{T_2^{\rm IP}} \ldots \int_{T_{nr}^{\rm N}}^{T_{n_r}^{\rm IP}} \frac{1}{T_1 T_2 \ldots T_{nr}} dT_1 dT_2 \ldots dT_{n_r}$$
$$= \frac{\tau B}{\ln 2} \ln \frac{T_1^{\rm IP}}{T_1^{\rm N}} \ln \frac{T_2^{\rm IP}}{T_2^{\rm N}} \ldots \ln \frac{T_{n_r}^{\rm IP}}{T_{nr}^{\rm N}} \qquad (19)$$

(19)式中：$T_j^{\rm IP} = \dfrac{U_j + U_j^{\rm FEC} + U_j^{\rm N}}{k_{\rm B} \ M_j^{\rm S} + M_j^{\rm FEC} + M_j^{\rm N} \ \ln 2}$。我们将信道容量 $C$ 定义为单位时间内接收到的总信息量 $bit/s$：

$$C = \frac{\sum_{j=1}^{n_{\rm r}} M_j^{\rm O}}{\tau} \qquad (20)$$

(20)式中：$\sum_{j=1}^{n_{\rm r}} M_j^{\rm O} = \int_0^{\sum_{j=1}^{n_{\rm r}} M_j^{\rm O}} dM^{\rm O}$，结合(19)式，最终我们得到了一个基于热力学的广义信道容量

$$C = B \times \sum_{i=1}^{\min \ n_{\rm t}, n_{\rm r}} \left( \log_2 \left( 1 + \frac{S_i + P_i^{\rm FEC}}{N_i} \right) - \log_2 \left( 1 + \frac{M_i^{\rm S} + M_i^{\rm FEC}}{M_i^{\rm N}} \right) \right) \qquad (21)$$

(21)式中：$S_i$、$P_i^{\rm FEC}$、$N_i$ 分别是第 $i$ 条发送支路的信号功率、前向纠错码功率和噪声功率，功率 $P$ 与能量 $U$ 的关系为 $U = P\tau$。(21)式显示出我们提出的基于热力学的广义信道容量公式与 MIMO 系统香农信道容量之间的一致性：$m \times m$ 的 MIMO 系统信道容量的极限，不会大于 $m$ 个独立的 SISO 子信道容量极限之和。第一项表明信道容量可以被重新表述为热力学定律的结果，印证了热力学与通信之间存在的内在紧密关联。(21)式中由自由度表征的第二项表明，在现实通信场景下的信号自由度和噪声自由度对信息传输速率的约束作用。进一步地，由对数函数的性质我们给出热力学信道容量的上下界，如(22)式所示。

$$C_{\rm LO} \le C \le C_{\rm HI} \qquad (22)$$

$$C_{\rm LO} = \log_2 \left( 1 + \prod_{i=1}^{n} \frac{S_i + P_i^{\rm FEC}}{N_i} \right) - \min \ n_{\rm t}, n_{\rm r} \ \times \log_2 \left( 1 + \frac{\sum_{i=1}^{\min \ n_{\rm t}, n_{\rm r}} \left( \frac{M_i^{\rm S} + M_i^{\rm FEC}}{M_i^{\rm N}} \right)}{\min \ n_{\rm t}, n_{\rm r}} \right)$$

$$C_{\rm HI} = \min \ n_{\rm t}, n_{\rm r} \ \times \log_2 \left( 1 + \frac{\sum_{i=1}^{\min \ n_{\rm t}, n_{\rm r}} \left( \frac{S_i + P_i^{\rm FEC}}{N_i} \right)}{\min \ n_{\rm t}, n_{\rm r}} \right) - \log_2 \left( 1 + \prod_{i=1}^{\min \ n_{\rm t}, n_{\rm r}} \ \frac{M_i^{\rm S} + M_i^{\rm FEC}}{M_i^{\rm N}} \right)$$

## 3　仿真分析

部分仿真参数设置见表 1。

**表 1　仿真的主要参数**

Table 1. Main parameters of the modulation

| 仿真参数 | 数值 |
|---|---|
| 调制方式 | 64QAM |
| 发端天线数 $n_{\rm t}$ | 128 |
| 收端天线数 $n_{\rm r}$ | 4 |
| 噪声温度 $T^{\rm N}$ | 298.15K（室温） |
| 带宽 $B$ | 20 MHz |
| 玻尔兹曼常数 $k_B$ | $1.38 \times 10^{-23}$ |

图 4 展示了热力学信道容量随噪声自由度变化的关系图。随着噪声自由度的增大，$M^{\rm N} \gg M^{\rm S} + M^{\rm FEC}$，热力学信道容量将越来越逼近香农信道容量，(21)式在这种情况下退化为(23)式。从图 4 中可以看出，相对来说，高斯噪声对热力学信道容量来说是“有益”的噪声，而脉冲噪声的影

响相比之下要“恶劣”得多,因为在这种情况下，噪声的自由度为 1。

$$C = B \times \sum_{i=1}^{\min n_t, n_r} \left( \log_2 \left( 1 + \frac{S_i + P_i^{FEC}}{N_i} \right) \right) \quad (23)$$

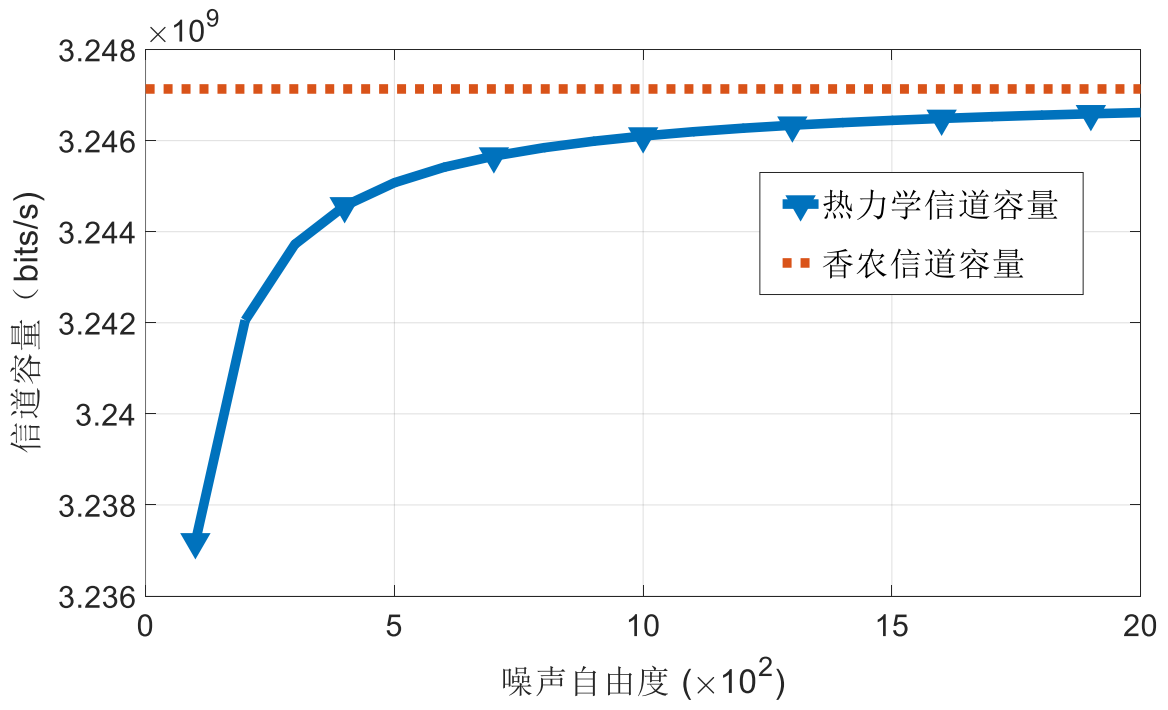

**图 4 热力学信道容量随噪声自由度变化的关系图**

Fig.4 The relationship between the channel capacity and the degrees of freedom of noise

图 5 展示了编码开销 $\psi$ 和热力学信道容量 $C$ 之间的变化关系。编码开销 $\psi$ 定义为冗余码自由度与源信号自由度的比值，即 $\psi = M^{FEC} / M^{S}$ .由图 5（a）可以看到，随着编码开销的增加，热力学信道容量也在增加，而文中推导的热力学信道容量和香农容量之间有着一定的差距。这种差距从(21)式的第二项可以直观看出，是发送信号、前向纠错码和噪声存在的自由度导致的。图 5（b）中展示了基于热力学推导的信道容量上下界的变化趋势，热力学信道容量上下界由公式(22)推导而得，阴影区域表示热力学信道容量的取值范围。随着编码开销的增加，热力学信道容量的增加出现边际递减效应。

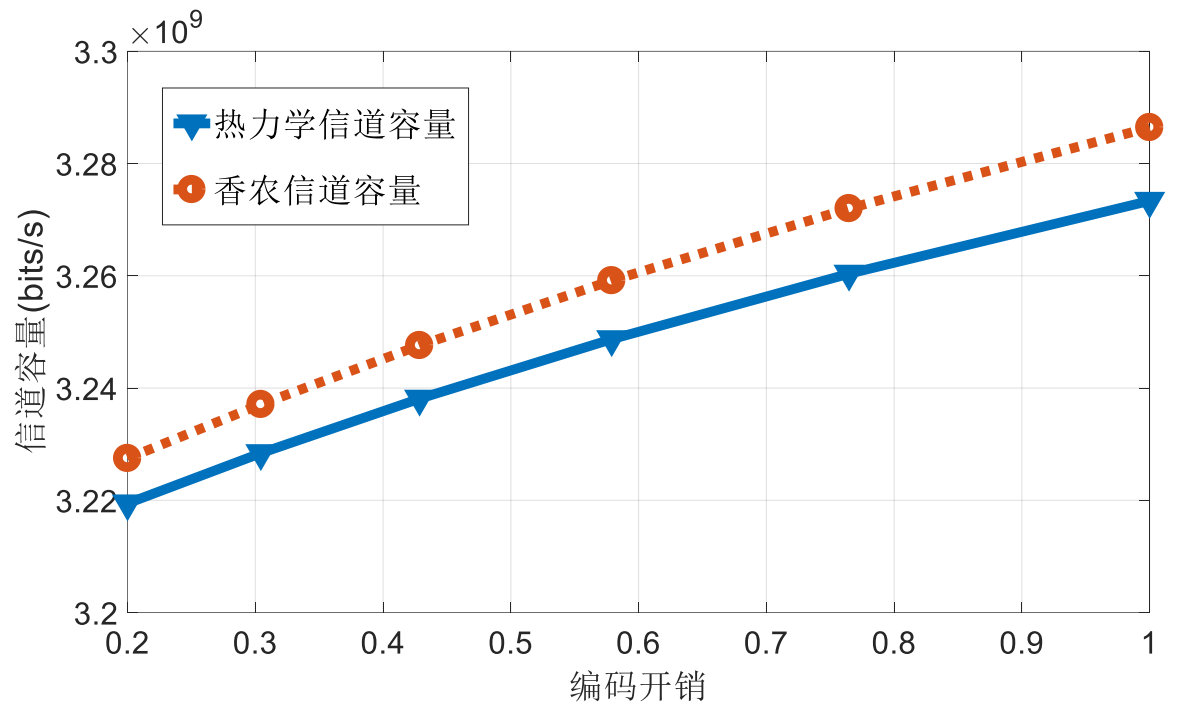

**(a) 理论香农容量和热力学信道容量的对比**

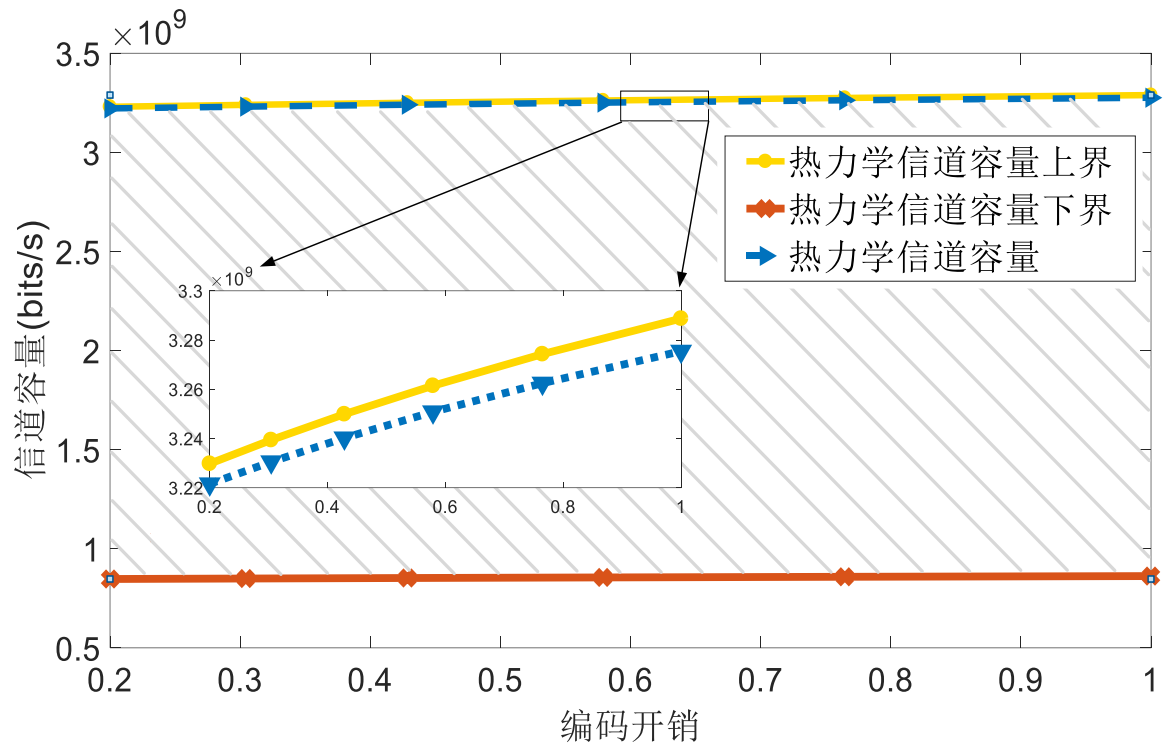

**（b）热力学信道容量上下界**

**图 5 热力学信道容量随编码开销变化的关系图**

Fig.5 Diagram of the relationship between channel capacity and coding overhead (a)Comparison of theoretical Shannon capacity and derived channel capacity (b)upper and lower bounds of channel capacity

## 4 结 论

热力学理论是分析能量流动的普遍适用的工具。与许多其他自然过程和系统类似，我们将通信系统和环境交互的过程看作是耗散转换过程，这是由于参与系统与其周围环境之间存在着能量的交互。本文介绍了如何基于热力学联合传统卡诺热机对大规模 MIMO 通信网络的信息流与能量流进行建模，引入热力学参量对通信系统各部分进行表征与模型推导，最终得到了基于热力学的广义信道容量表达式。仿真结果展示了热力学信道容量和香农容量之间的一定差距，并给出了解释。在未来 5G 通信能效优化场景中[22], [23]，结合热力学分析理论和传统信息论构建优化模型是一个有意义的开放性问题，而该问题的解决将推 动新一代通信技术革命。

## 参考文献

[1] Samardzija D. Some analogies between thermodynamics and Shannon theory[C]//2007 41st Annual Conference on Information Sciences and Systems. IEEE, 2007: 166-171.

[2] Perabathini B, Varma V S, Debbah M, et al. Physical limits of point-to-point communication systems[C]//2014 12th International Symposium on Modeling and Optimization in Mobile, Ad Hoc, and

Wireless Networks (WiOpt). IEEE, 2014: 604-610.

[3] Ganesh N, Anderson N G. Dissipation in neuromorphic computing: Fundamental bounds for feedforward networks[C]//2017 IEEE 17th International Conference on Nanotechnology (IEEE-NANO). IEEE, 2017: 594-599.

[4] Kostina V, Polyanskiy Y, Verdú S. Transmitting k samples over the Gaussian channel: Energy-distortion tradeoff[C]//2015 IEEE Information Theory Workshop (ITW). IEEE, 2015: 1-5.

[5] Sourlas, N. Spin-glass models as error-correcting codes[J].Nature 339, 693 - 695(1989).

[6] H. Nishimori, Statistical physics of spin glasses and information processing: an introduction[M]. Clarendon Press, 2001.

[7] O. Shental and I. Kanter. Shannon Meets Carnot: Mutual Information Via Thermodynamics[J]. arXiv preprint arXiv:0806.3133, 2008.

[8] S. Deffner and C. Jarzynski. Information Processing and the Second Law of Thermodynamics: An Inclusive, Hamiltonian Approach[J]. Physical Review X, 2013, 3(4): 041003.

[9] I. Kanter, O. Shental, H. Efraim, and N. Yacov. Carnot in the Information Age: Discrete Symmetric Channels[EB/OL]. (2008-6)[2021-10-29]. . http://arxiv.org/abs/0807.4322

[10] Parker M C, Walker S D. A unified carnot thermodynamic and shannon channel capacity information-theoretic energy efficiency analysis[J]. IEEE Transactions on Communications, 2014, 62(10): 3552-3559.

[11]Chen L, Sun F, Wu C. Effect of heat transfer law on the performance of a generalized irreversible Carnot engine[J]. Journal of Physics D: Applied Physics, 1999, 32(2): 99.

[12]Ge X, Sun Y, Gharavi H, et al. Joint optimization of computation and communication power in multi-user massive MIMO systems[J]. IEEE transactions on wireless communications, 2018, 17(6): 4051-4063.

[13]Ge X, Yang B, Ye J, et al. Spatial spectrum and energy efficiency of random cellular networks[J]. IEEE Transactions on Communications, 2015, 63(3): 1019-1030.

[14]Aleksic S. Energy, entropy and exergy in communication networks[J]. Entropy, 2013, 15(10): 4484-4503.

[15]Zenil H, Kiani N A, Tegnér J. The thermodynamics of network coding, and an algorithmic refinement of the principle of maximum entropy[J]. Entropy, 2019, 21(6): 560.

[16]Kafri O. Informatics Carnot Machine[J]. arXiv preprint arXiv:0705.2535, 2007.

[17]Brissaud J B. The meanings of entropy[J]. Entropy, 2005, 7(1): 68-96.

[18]Keizer J. Heat, work, and the thermodynamic temperature at nonequilibrium steady states[J]. The Journal of chemical physics, 1985, 82(6): 2751-2771.

[19]Landauer R. Irreversibility and heat generation in the computing process[J]. IBM journal of research and development, 1961, 5(3): 183-191.

[20]Ganesh N, Anderson N G. On-chip error correction with unreliable decoders: Fundamental physical limits[C]//2013 IEEE International Symposium on Defect and Fault Tolerance in VLSI and Nanotechnology Systems (DFTS). IEEE, 2013: 284-289.

[21]Landauer R. Dissipation and noise immunity in computation and communication[J]. Nature, 1988, 335(6193): 779-784.

[22]Ge X, Ye J, Yang Y, et al. User mobility evaluation for 5G small cell networks based on individual mobility model[J]. IEEE Journal on Selected Areas in Communications, 2016, 34(3): 528-541.

[23]Zhong Y, Quek T Q S, Ge X. Heterogeneous cellular networks with spatio-temporal traffic: Delay analysis and scheduling[J]. IEEE Journal on Selected Areas in Communications, 2017, 35(6): 1373-1386.

## 附录

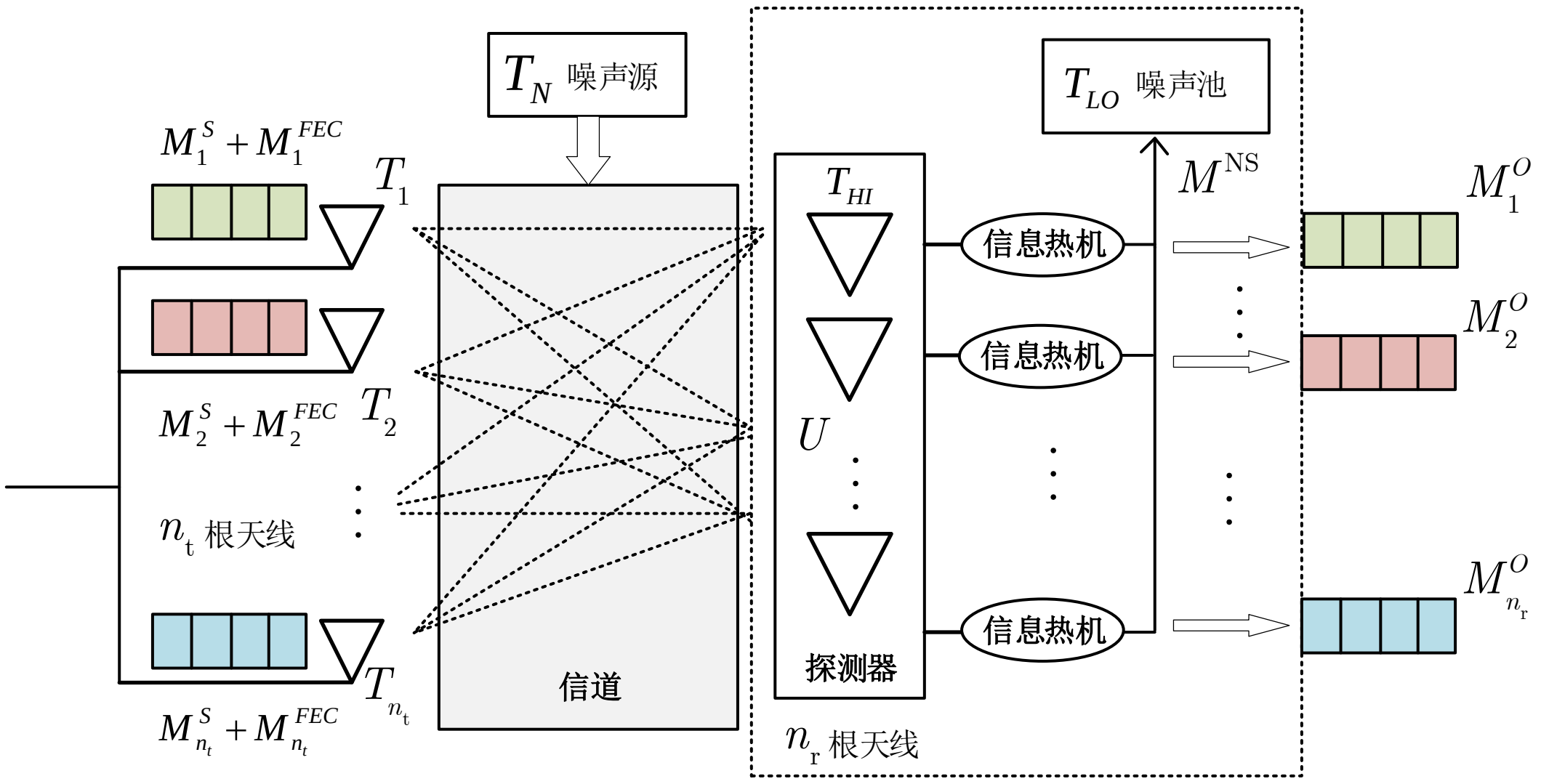

**图 2 MIMO 通信系统的热力学模型示意图**

Fig.2 Schematic diagram of thermodynamic model of MIMO communication system

## [作者简介]

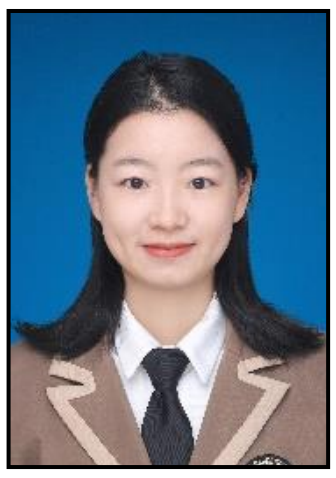

阮慕瑶（1998- ），女，湖北咸宁人，硕士，华中科技大学研究生，主要研究方向为信息热力学，无线通信等。

E-mail：1029282394@qq.com

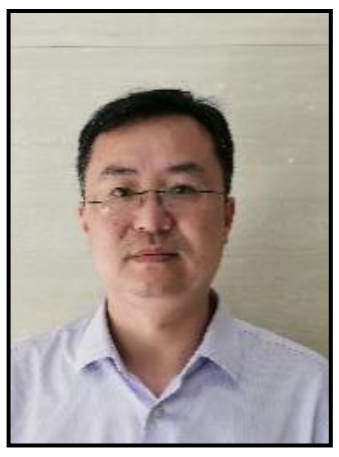

葛晓虎（1972− ），男，湖北武汉人，博士，华中科技大学教授，主要研究方向为移动通信、无线网络中的流量建模、绿色通信等。
E-mail：xhge@hust.edu.cn